\documentclass[pra,aps,showpacs,twocolumn]{revtex4}
\usepackage{amssymb}
\usepackage{times}
\usepackage{amsmath}
\usepackage{graphicx}
\usepackage{dcolumn}
\usepackage{epsfig}
\usepackage{bm}
\usepackage{braket}
\usepackage{mathrsfs}
\usepackage{hyperref}
\begin{document}
\title{{\Large{Microwave controlled efficient Raman and sub-Raman generation}}}
\author{Pankaj K. Jha$^{1,}$\footnote{Present Address: University of California, Berkeley, California 94720, USA. Email: pkjha@berkeley.edu}, Sumanta Das$^{2,}$\footnote{Currently at Max-Planck Institut f\"{u}r Kerphysik, 69117 Heideberg, Germany. Email: sumanta.das@mpi-hd.mpg.de}, Tarak N. Dey$^{3,}$\footnote{Email: tarak.dey@iitg.ernet.in}}
\affiliation{$^{1}$Texas A$\&$M University, College Station, Texas-77840,  USA\\
$^{2}$Oklahoma State University, Stillwater, Oklahoma-74078, USA\\
$^{3}$Indian Institute of Technology Guwahati, Guwahati-781039, Assam, India}
\date{\today}
\begin{abstract}
We propose an efficient scheme for the generation and the manipulation of Raman fields in an homogeneously broadened atomic vapor in a closed three levels $\Lambda$-configuration. The key concept in generating the Raman and sub-Raman fields efficiently at lower optical densities involve the microwave induced atomic coherence of the lower levels. We show explicitly that, generation efficiency of the Raman fields can be controlled by manipulating the coherences via phase and amplitude of the microwave field.
\end{abstract}
\maketitle
\section{Introduction}
Coherence effects in a multilevel atomic system induced by coherent electromagnetic fields has attracted considerable attention due to intriguing counterintuitive physics and potential important applications\cite{MOS}. Induced atomic coherences display many optical phenomena such as enhanced nonlinear effects\cite{Tewari_PRL_86}, electromagnetically induced transparency (EIT)\cite{Harris90, Boller91,Field91}, giant Kerr nonlinearity\cite{Schmidt99,Harris99,Wang01},  lasing  without inversion\cite{LWI1,LWI2,LWI3}, efficient nonlinear frequency conversions \cite{Jain96,Jain93,Xiao96}, relative intensity squeezing\cite{Arimondo_OL_07}, coherence Raman scattering enhancement via maximum coherence in atoms\cite{Merriam00,Scully02} and molecules\cite{Sari04,Bead05,Pestov07}, enhanced lasing\cite{Jha12, Thesis12, Sete12}, coherent Raman umklappscattering\cite{CRU}, controlled resonance profiles\cite{Dorf11}, carrier-envelope phase effects by multi-cycle pulses\cite{Jha1} etc. to name a few.

Usually in such systems, the excitation of atomic or molecular coherence is achieved by optical fields. Generating and harnessing new coherences in the presence of  a microwave field, in addition to the optical fields in this systems have gain much attention is the past decade\cite{Wilson05,Kosa00}. For example, it was shown that microwave field can be utilized to envision novel effects like electromagnetically induced transparency with amplification(EITA)\cite{Joo10} in superconducting qubits, sub-wavelength atom localization\cite{Sahrai05}, circuit QED\cite{Murali04}, simultaneous slow and fast light\cite{Luo10}, gains without inversion in quantum systems with broken parities\cite{Jia10}, quantum storage\cite{Novikova07,Novikova07b,Gorshkov07} with features like minimal loss and distortion of the pulse shape. Recently Jha and Ooi studied the control of quantum resonances in photonic crystals with electromagnetically induced transparency driven by microwave field\cite{JhaAPB} and showed that the intensity and phase of the microwave field can alter the transmission and reflection spectra in interesting ways, producing hyperfine resonances.  Such ideas has opened the door for nanophotonic devices to be merged with microwave telecommunication devices
\begin{figure}[b]
 \label{Fig1}
\begin{center}
\centerline{\includegraphics[height=6.5cm,width=0.49\textwidth,angle=0]{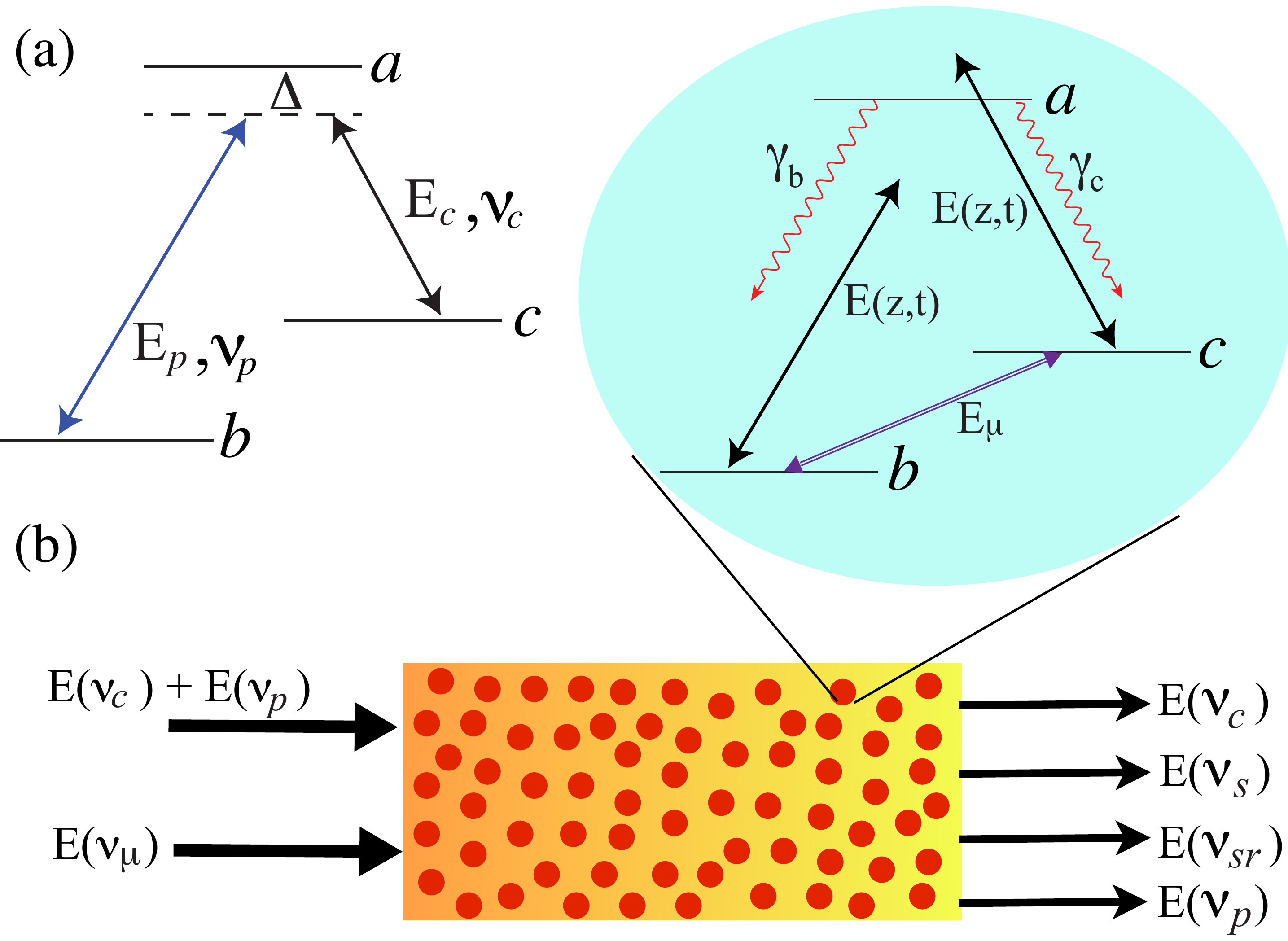}}  
\caption{(Color online) (a) Schematic diagram of a three-level atomic system with energy spacing $\omega_{cb}$ between two ground states $c$ and $b$ in a typical EIT setup. The control field with frequency $\nu_{c}$ and probe field with frequency $\nu_{p}$ act on the atomic transitions $|a\rangle \leftrightarrow |c\rangle $ and $|a\rangle  \leftrightarrow |b\rangle $, respectively. (b) Same three-level $\Lambda$ system as (a) but at high atomic number density and in presence of the microwave field. The space-time-dependent field $E(z , t)$ can now coupled to both the optical transitions generating Raman stokes and anti-stokes frequencies along with the pump and probe frequencies.}
\end{center}
\end{figure}

Typically in the $\Lambda$-configuration we use two lasers,  where one is use to excite the coherence (also known as the drive field) and the other (weak) is used as a probe. Here the underlying assumption is that the drive field does not couple the probe transition and vice-versa. This is a good approximation when the two frequencies are not close to each other or such coupling is not allowed by selection rules\cite{MOS}. Not so long ago Harada \textit{et al.}\cite{Harada06} investigated an interesting problem of competition between stimulated Raman scattering(SRS) and EIT in an atomic gas in $\Lambda-$configuration and showed that at higher optical density the conversion of probe photon to Stokes photon is not negligible. Later Agarwal \textit{et al.}\cite{Agarwal06} approached this problem theoretically and showed that at higher densities the pump photon is depleted due the multiple Raman scattering processes. In both the papers\cite{Harada06,Agarwal06} the optical density seems to be the controlling parameter of the two competing processes. 

In this paper, we report our theoretical study of efficient Raman and sub-Raman fields generation from homogeneously broadened and low density atomic vapor of Potassium. Here we model the Raman fields generation by atoms in $\Lambda-$configuration excited by optical and microwave fields. The optical fields couple the dipole allowed transition while the dipole forbidden transition is excited by the microwave field. Similar configuration has also been studied earlier in efficient FWM process as a result of stimulated Raman scattering of the optical field\cite{Zibrov02}.  Here we show that applying microwave fields, which induces atomic Zeeman coherence,  leads to efficient Raman field generation even at lower optical density contrary to that reported in its absence\cite{Agarwal06}.  The main result of the paper are shown in Figs. 2 and 5 where we have plotted the spectral amplitudes of the Raman stokes and sub-Raman fields generated as a function of the optical density. The paper is organized as follows. In section II, we obtain the equation of motion for the density matrix elements $\varrho_{ij}$. In section III, we present the findings of our numerical simulation where we have studied the evolution of the Raman and sub-Raman fields as a function of the optical density. In section IV, we summarize and present possible applications of our results.  
\section{Theory}
Our model system consist of a cell filled with a low atomic number density $(\sim 10^{10}$atoms/cm$^{3})$ gas of three-level atoms with energy levels in $\Lambda$-configuration as depicted in Fig. 1.  Here we consider that an optical beam propagating through the cell along $z$ axis can couple to both the dipole allowed transitions $|a\rangle \leftrightarrow |b\rangle$ and $|a\rangle \leftrightarrow |c\rangle$ \cite{Agarwal06}. The dynamics of the atomic system thus becomes notably different from a typical EIT scheme which is well known in such three-level atomic configurations [see Fig. 1(a)], and other nonlinear processes starts to become significant. To account for different nonlinear processes in our model, following \cite{Agarwal06}, we write the input field in the form,
\begin{equation}
\label{3}
E(z,t) = \mathcal{E}(z,t)e^{-i\nu_{c}t}+\text{c.c},
\end{equation}
where $\mathcal{E}(z,t)$ denotes the net generated field. At the input face of the medium $\mathcal{E}(z,t)$ has two components to account for both control and probe fields: 
\begin{equation}
\label{3a}
\mathcal{E}(z,t)=\mathcal{E}_{c}(z)+\mathcal{E}_{p}(z)e^{i(\nu_{p}-\nu_{c})t} ,
\end{equation}
where $\mathcal{E}_{j}(z) = \mathcal{E}e^{ik_{j}z}$ is the position dependent amplitude of the control and probe with $\nu_{c}$ and $\nu_{p}$ being the respective frequencies. 

The cell is placed in a microwave cavity and a microwave field couple the dipole forbidden Raman transition $|b\rangle  \leftrightarrow |c\rangle $. This create coherence among the lower states which as we will see later are important for enhancement of Raman generation in the system. We consider the microwave field as,
\begin{equation}
E_{\mu}(z,t) = \mathcal{E}_{\mu}(z)\cos(\nu_{\mu}t+\phi_{\mu}),
\end{equation}
where $\mathcal{E}_{\mu}(z)$ position dependent amplitude of the microwave field inside the cavity, $\nu_{\mu}$ is the microwave frequency, and $\phi_{\mu}$ is the phase of the microwave field. Under the Raman-resonance condition $\nu_{p}-\nu_{c} = \omega_{bc}$, we expect $\mathcal{E}(z,t)$ to have the time structure 
\begin{equation}
\mathcal{E}(t) = \sum \mathcal{E}^{(n)}e^{-in\omega_{bc}t}
\end{equation}
Here $\mathcal{E}^{(\pm n)}$ gives the strength of the $n$th order Raman fields for $n > 1$. The first order processes gives $\mathcal{E}^{(-1)}$ as the strength of the stokes field and $\mathcal{E}^{(+1)}$ describes the change in the probe field. The Hamiltonian for our three level system in the dipole approximation can be written as,
\begin{equation}
\begin{split}
\mathscr{H} =\sum_{k}\hbar\omega_{k}|k\rangle\langle k|&-[\wp_{ab}E(z,t)|a\rangle\langle b|+\wp_{ac}E(z,t)|a\rangle\langle c|\\
&+\wp_{cb}E_{\mu}(z,t)|c\rangle\langle b|+\text{h.c}]
\end{split}
\end{equation}
where $k = (a,b,c)$ and $\wp_{\alpha\beta} = \langle\alpha|\hat{\wp}|\beta\rangle$ is the dipole moment of the corresponding transition ($|\alpha\rangle \leftrightarrow |\beta\rangle$). To investigate the space-time dependent dynamics of the atomic system in presence of relaxation processes we take a master equation approach. The corresponding quantum Louville equation of motion is given by,
\begin{equation}
\label{eq5}
\begin{split}
\partial_{t }\rho= -\frac{i}{\hbar }\left[ \mathscr{H},\rho \right]+\mathcal{L}\rho\\
\end{split}
\end{equation}
where $\mathcal{L}\rho$ is given by
\begin{equation}
\mathcal{L}\rho = -\sum_{k,j}\frac{\gamma_{kj}}{2}(\sigma^{+}_{k}\sigma^{-}_{j}\rho+\rho\sigma^{+}_{k}\sigma^{-}_{j}-2\sigma^{-}_{j}\rho\sigma^{+}_{j}).
\end{equation}
Here $\rho$ is the density operator of the system, $\mathcal{L}\rho$ is the Lindbald operator representing the relaxation processes, $\gamma_{kk}$ is the spontaneous emission rate, $\gamma_{kj}$ is the decoherence rate of the relevant coherences $\rho_{kj}$ and $\sigma_{j}^{\pm}$ the atomic raising and lowering operators defined as $\sigma^{+}_{b} =\left| b\rangle \langle a\right|$ and $\sigma^{+}_{c} =\left| c\rangle \langle a\right|$. 

The population and coherences can be evaluated by expanding the above equation in the basis $(a, b, c)$. In doing so we work in a frame rotating with frequency $\nu_{c}$ to eliminate the highly oscillating terms and thus apply the transformations 
\begin{equation}
\rho _{ab}=\varrho _{ab}\exp [i(k_{c}z-\nu _{c}t)],
\end{equation}
\begin{equation}
\rho _{ac}=\varrho _{ac}\exp [i(k_{c}z-\nu_{c}t)],
\end{equation}
and $\rho_{cb}=\varrho_{cb}$. The density matrix equations governing the dynamics of the atomic coherences and populations with the above mentioned transformation is then given by, 
\begin{equation}\label{eqab}
\dot{\varrho}_{ab}= -\Gamma_{ab}\varrho_{ab}-i\Omega(\rho_{aa}-\rho_{bb})+i\Omega\varrho_{cb} -i\Omega_{\mu}e^{i\theta_{\mu}}\varrho_{ac},
\end{equation}
\begin{equation}
\dot{\varrho}_{ac}= -\Gamma_{ac}\varrho_{ac}-i\Omega(\rho_{aa}-\rho_{cc})+i\Omega\varrho^{\ast}_{cb} -i\Omega^{\ast}_{\mu}e^{-i\theta_{\mu}}\varrho_{ab},
\end{equation}
\begin{equation}
\dot{\varrho}_{cb}=-\Gamma_{cb}\varrho_{cb}-i\Omega_{\mu}e^{i\theta_{\mu}}(\rho_{cc}-\rho_{bb})+i\Omega^{\ast}\varrho_{ac} -i\Omega\varrho^{\ast}_{ab},
\end{equation}
\begin{equation}
\dot{\varrho}_{aa}=-(\gamma_{b}+\gamma_{c})\varrho_{aa} +i\Omega\varrho_{ba}-i\Omega^{\ast}\varrho_{ab}+i\Omega\varrho_{ca}-i\Omega^{\ast}\varrho_{ac},
\end{equation}
\begin{equation}\label{eqbb}
\dot{\varrho}_{bb}=-\gamma_{b}\varrho_{bb} -i\Omega\varrho_{ba}+i\Omega^{\ast}\varrho_{ab}+i\Omega_{\mu}e^{i\theta_{\mu}}\varrho_{cb}-i\Omega^{\ast}e^{-i\theta_{\mu}}\varrho_{bc},
\end{equation}
where $\theta_{\mu} = k_{\mu}z-\nu_{\mu}t$ and the relaxation rates of the coherences are given by
\begin{equation}
\begin{split}
\Gamma_{ab} &= (\gamma_{a}+\gamma_{b})/2+\gamma_{p}+i\Delta',\\
\Gamma_{ac} &= (\gamma_{a}+\gamma_{c})/2+\gamma_{p}+i\Delta,\\
\Gamma_{cb} &= \gamma_{p}+i\omega_{cb}.
\end{split}
\end{equation}
where $\Delta'=\omega_{cb}-\Delta$ and $\Delta=\nu_{c}-\omega_{ac}$ . The spontaneous decay rate and coherence decay rates are represented by $\gamma_{k} (k = a,b,c)$ and $\gamma_{p}$ respectively.  The space-time dependent Rabi frequency corresponding to the optical and microwave fields are defined by $\Omega (z,t) = \vec{\wp}\cdot\vec{\mathcal{E}}_{\mu}/\hbar$ and $\Omega_{\mu}(z,t) = \vec{\wp}_{cb}\cdot\vec{\mathcal{E}}_{\mu}/\hbar$. For simplicity we have assumed the dipole moments of the optical transitions $\vec{\wp}_{ab} = \vec{\wp}_{ac} =\vec{\wp}$ and $\gamma_a=\gamma_b=\gamma_c=\gamma$. The equations for other coherences and populations easily follows from $\varrho_{jk} = \varrho_{kj}^{\ast}$ and $Tr(\varrho) = 1$.
\begin{figure}[t]
\label{Fig2}
\centerline{\includegraphics[height=8.0cm,width=0.49\textwidth,angle=0]{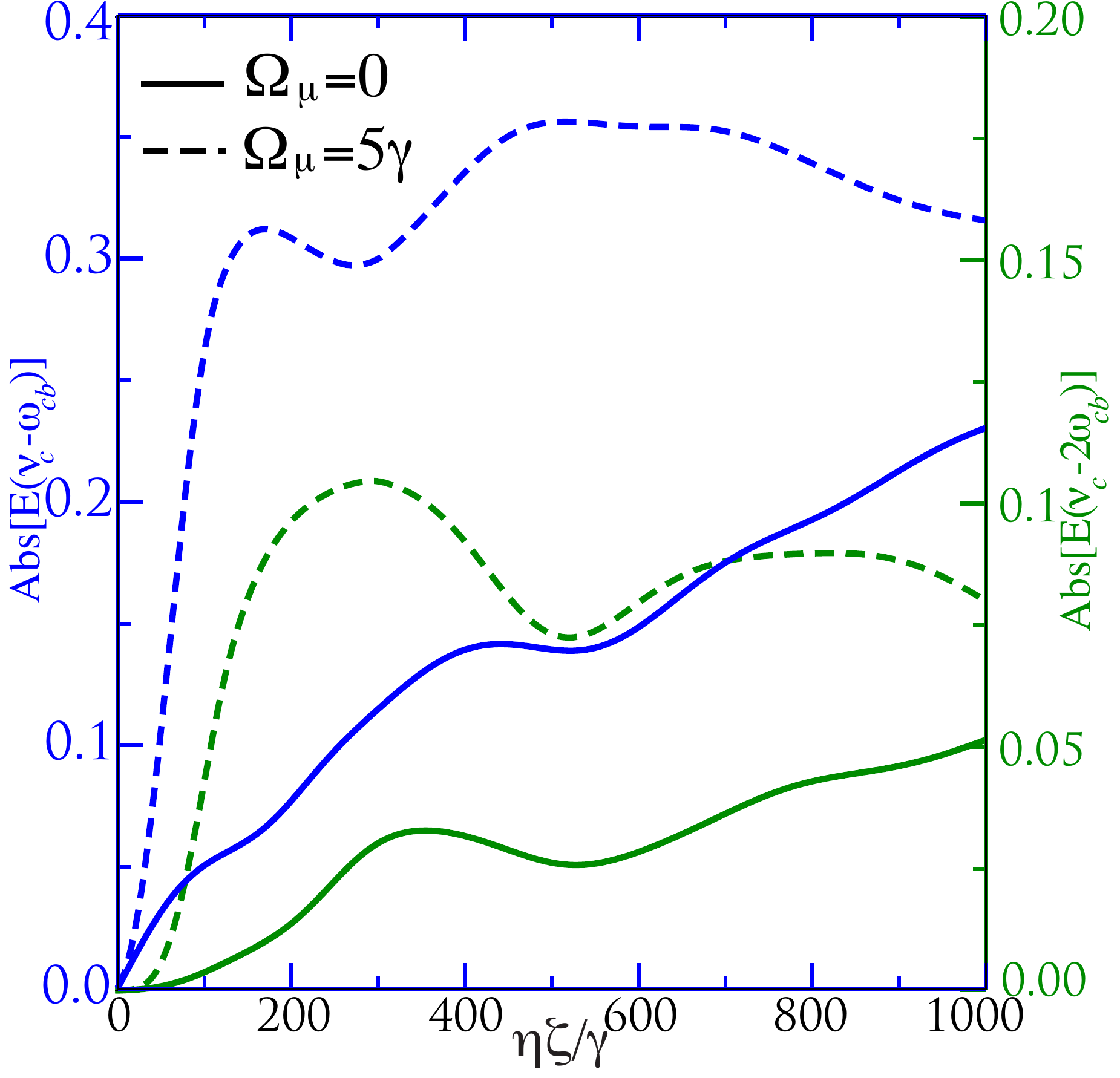}}
\caption{(Color online) (a) The spectral amplitudes of different Raman stokes and hyper-stokes fields are plotted as a function of the atomic density for a homogeneously broadened medium. The normalized propagation length $\eta\zeta/\gamma=200$ is equivalent to the actual length of the medium L=7.13 cm when the atomic density $n=2\times10^{10}$ atom/cm$^3$. For numerical simulation the common parameters of the curves for $^{39}$K vapor are chosen as: input control laser Rabi frequency $\Omega_{\cal E} =0.5\gamma$, ${\cal E}_c/{\cal E}_p=0.5$, $\Delta=50\gamma$, $\gamma_{p}=0.0$, $\omega_{cb}=100\gamma$ and 2$\gamma$=3.79$\times$10$^7$ rad/sec.}
\end{figure}
To investigate the non-linear response of the medium on the propagating optical field in presence of the microwave, we study the net generated field $\mathcal{E}(z,t)$ at the output for arbitrary atomic number density. The behavior of the generated fields is given by the Maxwell equations involving the induced macroscopic polarizations $\mathcal{P}(z,t)$, which in the slowly varying envelope approximation is 
\begin{equation}
\label{max}
 \frac{\partial E(z,t)}{\partial z}+\frac{1}{c}\frac{\partial E(z,t)}{\partial t}  = 2\pi i\kappa\mathcal{P}(z,t),
\end{equation}
The induced polarization is in turn connected to the coherences in the system by the relation, 
\begin{equation}
\vec{\mathcal{P}} = \left(\vec{\wp}_{ab}\rho_{ab}+ \vec{\wp}_{ac}\rho_{ac}\right)e^{-i\nu_{c}t}.
\end{equation}
\begin{figure}[t]
\label{Fig3}
\centerline{\includegraphics[height=7.2 cm,width=0.5\textwidth,angle=0]{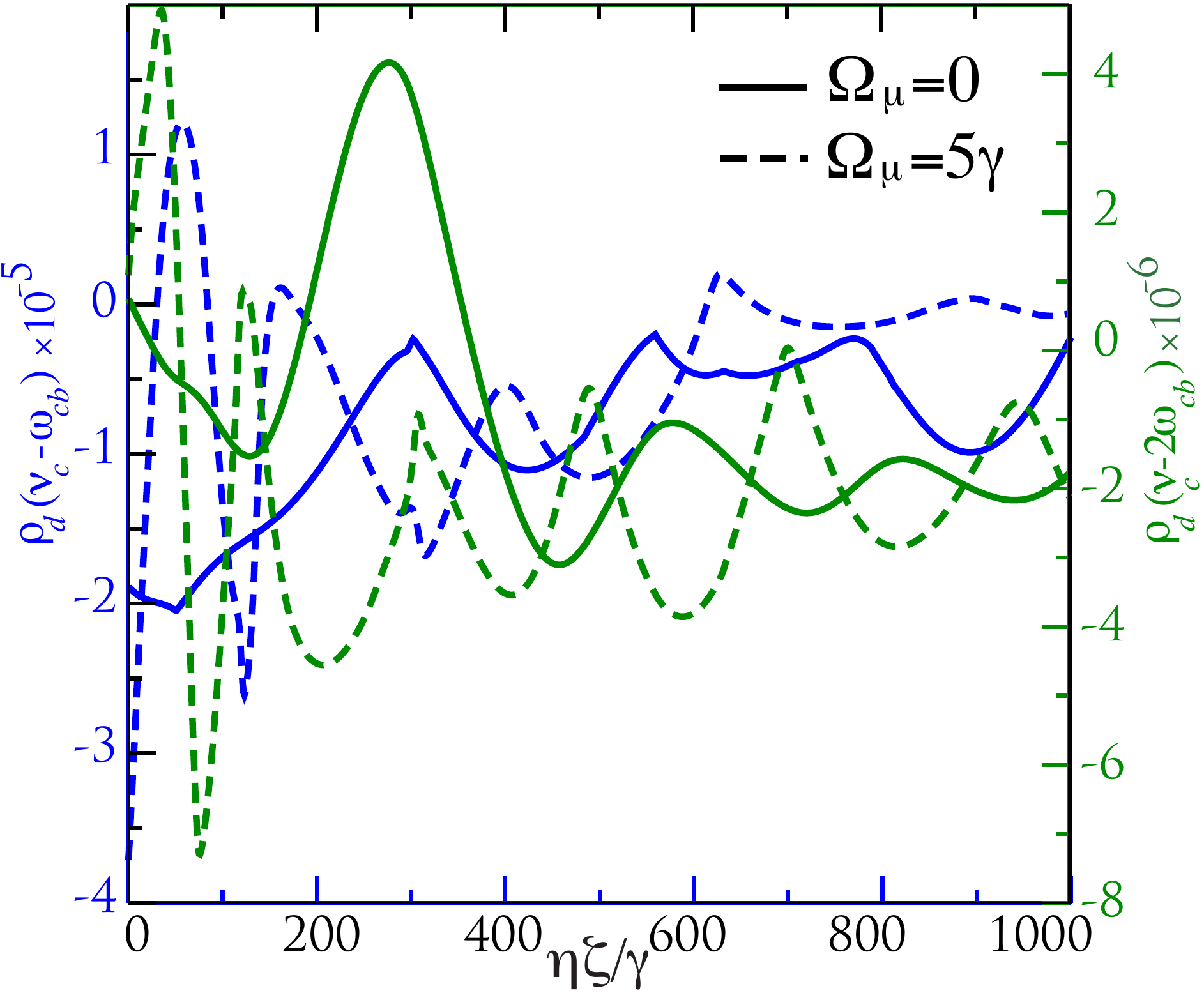}}  
\caption{(Color online) The spectral decomposition of the ground states population difference $\rho_d=\rho_{cc}-\rho_{bb}$ at the frequencies $ (\nu_c-\omega_{cb})$ and $ (\nu_c-2\omega_{cb})$ are plotted as a function of the atomic density. The used parameters are same as in Fig. 2}
\end{figure}
Thus Eq. (\ref{max}) then leads to a coupled Maxwell-Schrodinger equation for the generated field in the form
\begin{eqnarray}
\label{prop}
 \left( \frac{\partial\Omega}{\partial z}+\frac{1}{c}\frac{\partial\Omega}{\partial t} \right) = i\frac{\eta}{2}\left(\varrho_{ac}+\varrho_{ab}\right),
\end{eqnarray}
where $\eta = 3\lambda^{2}n\gamma/4\pi$. The coupled Eqs(\ref{eqab}-\ref{eqbb}) along with the propagation equation for the optical fields Eq.(\ref{prop}) are solved numerically in the moving co-ordinate system
\begin{equation}
\tau = t-z/c, \quad \zeta = z.
\end{equation}  
with the following initial conditions: atoms are in initially in the state $b$, microwave field is in resonance with the $b \leftrightarrow c$ transition, the medium is homogeneous and the fields at the input face of the medium is given by Eq.(\ref{3a}). We calculate the output field $\mathcal{E}(l,\tau)$ in terms of the space-time dependent Rabi frequency $\Omega(l,\tau)$ and following \cite{Agarwal06}, do a fast Fourier transform to obtain the different Fourier components of the field at the output face of the medium. This procedure enables us to determine the Raman signals and also provide us knowledge about the evolution of the probe and control fields. 
\section{Results and Discussion}
We first present the key findings of our work and discuss the consequences of the microwave field induced coherence of the lower levels on the generated output fields. Our results are based on simulation carried out with realistic experimental parameters corresponding to the level scheme $4S_{1/2}(F = 1) \equiv |b\rangle$, $4S_{1/2}(F = 2) \equiv |c\rangle$ and $4P_{1/2}\equiv |a\rangle$ of $^{39}K$ given in\cite{Conc97}. In Figs. 2 we show the generation of Raman stokes and hyper-stokes fields at the frequencies $(\nu_{c}-\omega_{bc})$ and $(\nu_{c}-2\omega_{bc})$ in the presence and the absence of the microwave coupling for low atomic number density $\sim10^{10}$.

The spectral amplitudes of Raman stokes and hyper-stokes fields show distinctly different behavior in the presence and the absence of the microwave fields. At optical densities $\sim 100$ there is substantial generation of the Raman fields in presence of the microwave. The generated signals are substantially stronger in comparison to the situation where the microwave coupling is absent. Note that some earlier work has shown generation of such Raman field in atomic gases but only at a much higher optical densities \cite{Harada06,Agarwal06}.  In Fig.2 we furthermore show in accordance to the earlier studies \cite{Harada06,Agarwal06} that in absence of the microwave coupling optical densities in excess of $1000$ is required to generate Raman fields of similar strength. 
\begin{figure}[t]
\label{Fig4}
\begin{center}
\centerline{\includegraphics[height=7.5cm,width=0.5\textwidth,angle=0]{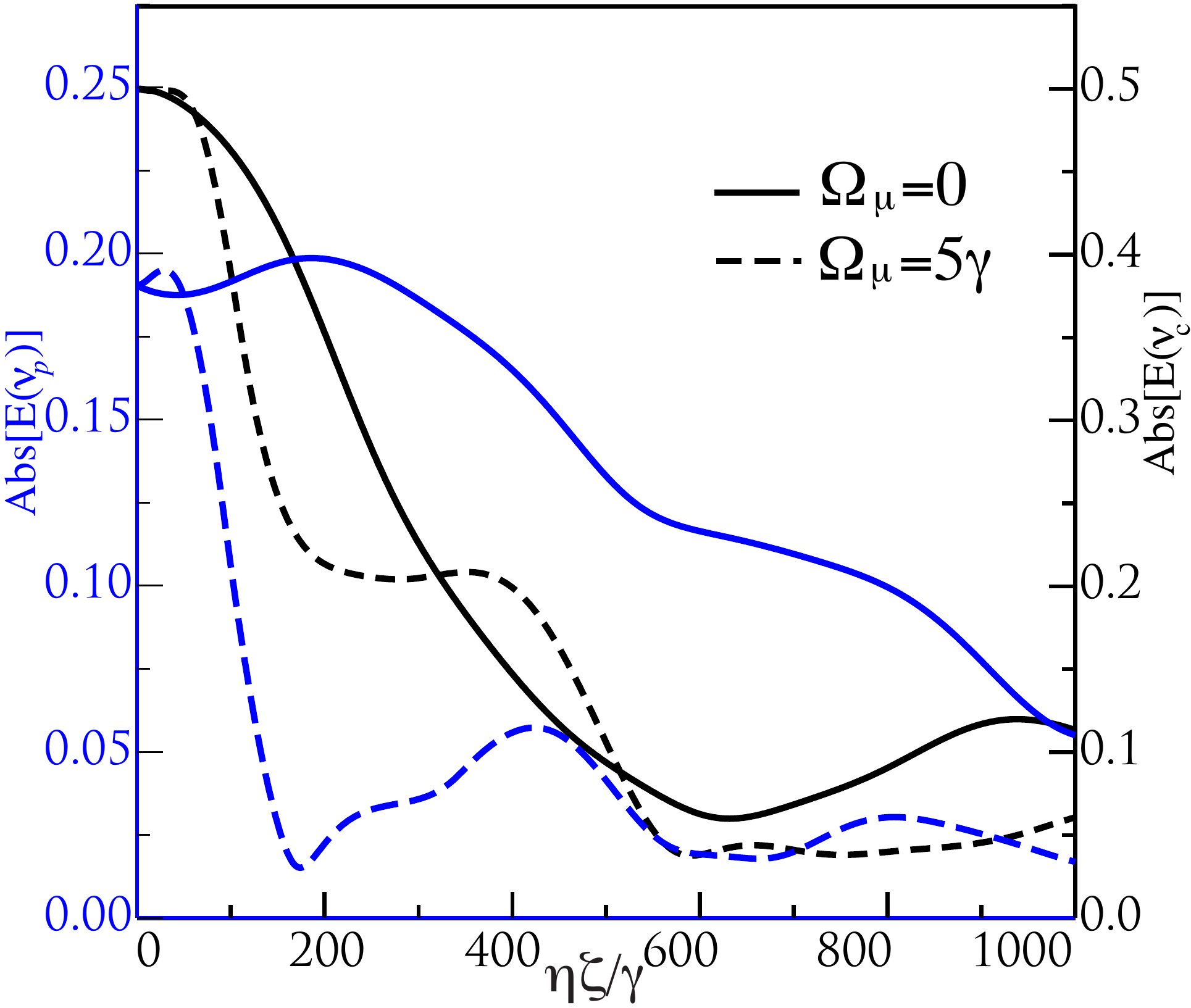}} 
\caption{(Color online) The spectral amplitudes of sub-Raman fields is plotted as a function of the atomic density for a homogeneously broadened medium. The parameters are same as Fig. 2.}
\end{center}
\end{figure}
\begin{figure}[t]
\label{Fig5}
\begin{center}
\centerline{\includegraphics[height=7.5cm,width=0.5\textwidth,angle=0]{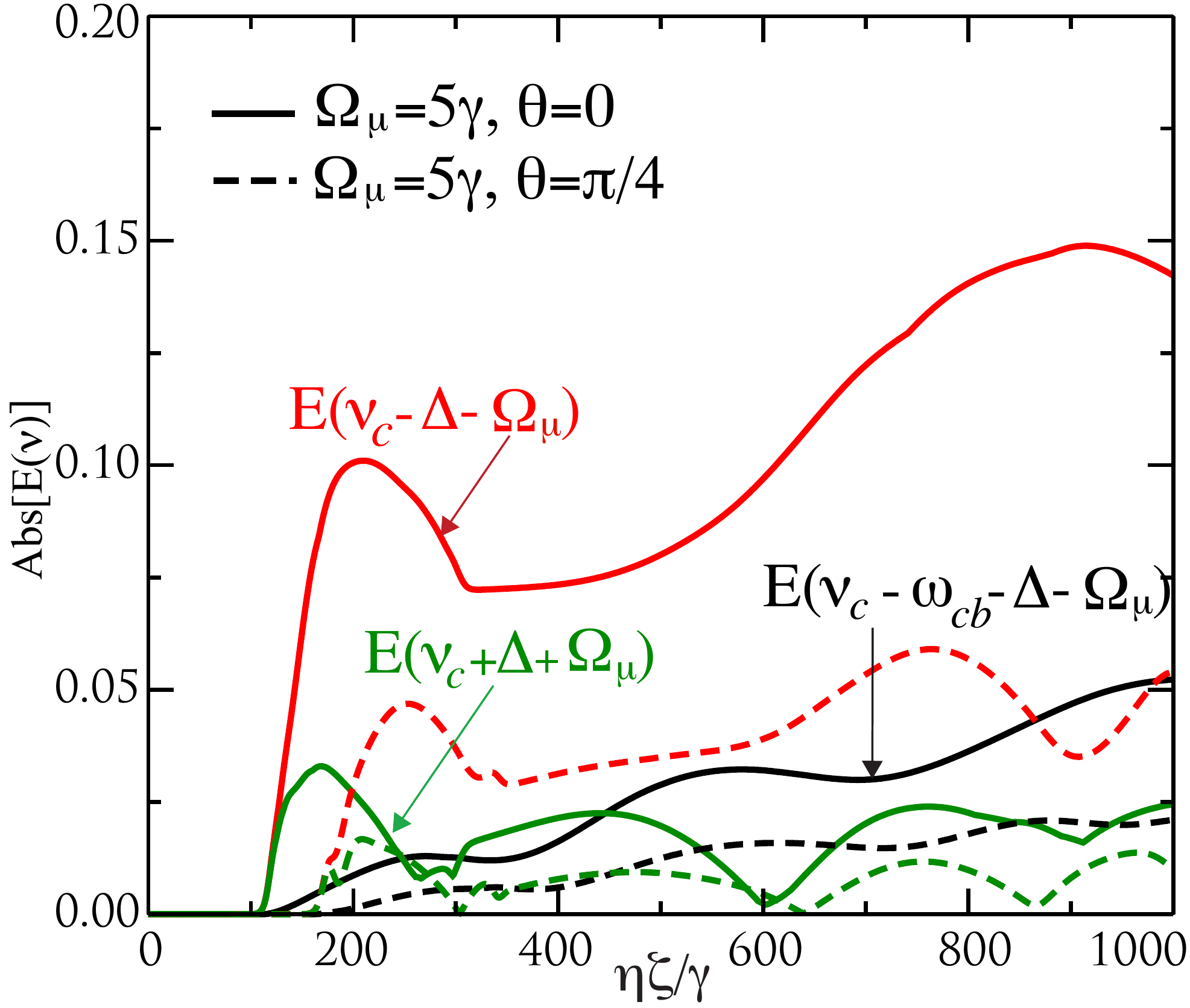}} 
\caption{(Color online) The spectral amplitudes of sub-Raman fields is plotted as a function of the atomic density for a homogeneously broadened medium. The parameters are same as Fig. 2.}
\end{center}
\end{figure}
These  Raman stokes and hyper stokes fields generation at low density regime can be explained by investigating the Fourier components of the population difference between the levels $|b\rangle$ and $|c\rangle$ at frequency $\nu_c-\omega_{cb}$ and $\nu_c-2\omega_{cb}$, respectively. It is clear from the Fig. 3 that  the population difference $\rho_d=\rho_{cc}-\rho_{bb}$ at the frequency $ (\nu_c-\omega_{cb})$ and $ (\nu_c-2\omega_{cb})$ differ significantly in presence and absence of the microwave field. As the cross-section for Raman is proportional to the population difference $\rho_d$ thus the microwave field induced population transfer at the lower state $|c\rangle$ enhances the Raman stokes and hyper stokes generation in even low optical densities in atomic gases. 

In Fig. 4 we show the behavior of the spectral amplitude of control and probe field at frequencies $\nu_{c}$ and $\nu_{p}$ respectively. We find that in the presence of the microwave coupling the probe field is strongly suppressed at optical density $\sim100$. In the absence of the microwave the probe field amplitude decreases gradually with an increase in the optical density. The control field amplitude on the other hand does not show much deviation in its behavior in presence or absence of microwave except for optical densities in certain range where the probe amplitude is seen to oscillate. Under this situation the control field amplitude falls. In general, all the components are coupled to each other via the Maxwell-Bloch equations and evolve dynamically as shown in Fig. 2 and Fig. 4.

In Fig. 5 we show the behavior of spectral amplitudes of sub-Raman fields as a function of the optical density in presence of microwave field with and without a phase. Here $\Delta=50\gamma$ is the detuning of the control field with respect to the transition $|a\rangle\leftrightarrow |c\rangle$. We find that for the phase $\theta=\pi/4$ the generation of sub-Raman fields are suppressed at low optical densities, while for $\theta=0$ we see strong sub-Raman generation even at lower optical densities around 100. We also noticed that complete suppression of the sub-Raman fields can be achieved at $\theta=\pi/2$ which is not shown in Fig. 5. Hence the phase of the microwave field also plays a important role in the generation of sub-Raman fields. This clearly manifest that the strength of coherence of the lower levels that is effective in Raman generation can be manipu- lated using the phase of the microwave field. Note that such phases have been shown to be useful in manipulation of EIT in atomic gases \cite{Li09}. However the strength of the signal varies according to the frequencies due to the phase dependent microwave induced population transfer from the level $|b\rangle \rightarrow |c\rangle$ or vice versa. Therefore, the sub Raman stokes lines are seen to be enhanced much more compare to the sub Raman anti-stokes. We would like to emphasize here that our numerical simulation suggest that this phase dependence is a generic behavior that occurs for different multiples of $n$. For discussion purpose however we have provided results for only one such phase in the manuscript. As such $\theta=0, \pi/4$ can be generalized to the form $\theta=n\pi, n=0, 1, 2, ...$ and $\theta=(n+1/2)\pi/2, n=0, 1, 2, ...$ respectively.

\section{Conclusion}
To conclude, we have investigated the efficient generation and manipulation of Raman fields by microwave induced coherence of the lower levels in a typical $\Lambda$ system.  Our proposed scheme works at remarkable low atomic densities. This is contrary to earlier studies in $\Lambda$ system \cite{Harada06, Agarwal06} where higher optical densities (almost $10$ times of what we consider) were required for Raman generation in absence of the microwave coupling.  Furthermore we also demonstrate that the phase of the microwave field can also be used as an additional control parameter in generation of the sub-Raman field in such $\Lambda$ system. Our numerical simulation were based on realistic experimental parameters and can thus be implemented for efficient Raman field generation in atomic gases. With recent demonstration of the technique called Laser induced atomic desorption(LIAD)\cite{Gozzini, Mariotti} combined with Raman scattering in the backward direction, demonstrated by Jha $et\, al.$\cite{JhaAPL12} in cesium vapor, this microwave controlled efficient Raman generation can be used to envision remote detection of chemicals at low vapor concentration. Furthermore, at low atomic densities our technique can be useful to manipulate the probe absorption in EIT medium\cite{Li09} by using the microwave field as the control knob. Extension of this work beyond atomic vapors to photonic crystals or superstructures will help in making a significant step in the direction of bridging nano-photonics with microwave devices and promoting the practical use of quantum coherence in metamaterials and photonic crystals to a wider domain.

\acknowledgments
One of the authors(T.N.D) gratefully acknowledges funding by the Science and Engineering Research Board(SR/S2/LOP-0033/2010). P. K. Jha is supported by the Herman F. Heep and Minnie Belle Heep Texas A\&M University Endowed Fund held/administered by the Texas A\&M Foundation, the Welch Foundation and the Texas A\&M Graduate Student Research and Presentation Grant.

\end{document}